\documentclass{article}
\usepackage{amsmath}

\newtheorem{theorem}{Theorem}

\newtheorem{definition}[theorem]{Definition}

\newtheorem{lemma}[theorem]{Lemma}

\newenvironment{proof}[1][Proof]{\noindent\textbf{#1.} }{\ \rule{0.5em}{0.5em}}
\input{tcilatex}

\begin{document}

\title{Quantum Probability from Decision Theory?}
\author{Simon Saunders}

\begin{center}
{\Huge Derivation of the Born Rule from \bigskip }

{\Huge Operational Assumptions}

\bigskip

{\LARGE Simon Saunders\footnote{%
Contact: simon.saunders@linacre.ox.ac.uk, and http://users.ox.ac.uk/\symbol{%
126}ppox/}}
\end{center}

\section{Introduction}

\noindent Whence the Born rule? It is fundamental to quantum mechanics; it
is the essential link between probability and a formalism which is otherwise
deterministic; it encapsulates the measurement postulates. Gleason's theorem 
\cite{Gleason} is mathematically informative, but its premises are too
strong to have any direct operational meaning: here the Born rule \ is
derived more simply, from purely operational assumptions.

The argument we shall present is based on Deutsch's derivation of the Born
rule from decision theory \cite{Deutsch}. The latter was criticized by
Barnum \textit{et al }\cite{Barnum}, but their objections hinged on
ambiguities in Deutsch's notation that have recently been resolved by
Wallace \cite{Wallace}; here we follow Wallace's formulation. The argument
is not quite the same as Wallace's, however. Wallace draws heavily on the
Everett interpretation, as well as on decision theory; like Deutsch, he is
concerned with constraints on\ \textit{subjective} probability, rather than
any objective counterpart to it. In contrast, the derivation of the Born
rule that we shall present is independent of decision theory, independent of
the interpretation of probability, and independent of any assumptions about
the measuring process. As such it applies to all the major foundational
approaches to quantum mechanics.

We assume the conventional scheme for the description of experiments: an
initial state, measured observable, and set of macroscopic outcomes. Given a
description of this form, we assume there is a general algorithm for the
expectation value of the observable outcomes (the Born rule is such an
algorithm). The argument then takes the following form: for a particular
class of experiments there are definite rules for determining such
descriptions, based on simple operational rules, and theoretical assumptions
that concern only the state-preparation device, not the measurement device.
These rules imply that in general such experiments can be described in 
\textit{different} ways. But the algorithm we are looking for concerns the
expectation value of the observed outcomes, so applied to these different
descriptions, it must yield the \textit{same} expectation value. Constraints
of this form are in fact sufficient to force the Born rule. \textit{If }%
there is to be such an algorithm, then it is the Born rule.

\section{Multiple-Channel Experiments}

The kinds of experiments we shall consider are limited in the following
respects: they are repeatable; there is a clear distinction between the
state preparation device and the detection and registration device; and -
this the most important limitation - we assume that for a given
state-preparation device, preparing the system to be measured in a definite
initial state, the state can be resolved into \textit{channels,} each of
which can be independently blocked, in such a way that when only one channel
is open the outcome of the experiment is \textit{deterministic} - in the
sense that if there is any outcome at all (on repetition of the experiment)
it is always the \textit{same} outcome. We further suppose that for every
outcome there \ is at least one channel for which it is deterministic, and -
in order to associate a definite initial state with a particular region of
the apparatus - we suppose that all the channels are recombined prior to the
measurement process proper.

For an example of such an experiment that measures spin, consider a neutron
interferometer, where orthogonal states of spin (with respect to a given
axis) are produced \ by a beam-splitter, each propagating along different
arms of the interferometer, before being recombined prior to the measurement
of that component of spin. For an example that measures position, consider
an optical two-slit experiment, adapted so that the lensing system after the
slits first brings the light into coincidence, but then focuses it on
detectors in such a way that each can receive light from only one of the
slits. It is not too hard to specify an analogous procedure in the case of
momentum;\footnote{%
The conventional method for preparing a beam of charged partciles of
definite momentum (by selecting for deflection in a magnetic field) can be
adapted quite simply.} any number of familiar experiments can be converted
into an experiment of this kind.

We introduce the following notation. Let there be $d$ channels in all, with $%
D\leq d$ possible outcomes $u_{j}\in U$, $j=1,...,D$. These outcomes are
macroscopic events (e.g. positions of pointers). Let $M$ denote the
experiment that is performed when all the channels are open, and $M_{k},$ $%
k=1,...,d$ the (deterministic) experiment that is performed when only the $%
k^{\text{th}}$ channel is open. Let there be identifiable regions $%
r_{1},r_{2},...$ of \ the \ state-preparation device through which the
system to be measured must pass (if it is to be subsequently detected at all
- regardless of which channels are open). Call an experiment satisfying
these specifications a \textit{multiple-channel experiment. }

One could go further, and provide operational definitions of the initial
states in each case, but we are looking for a probability algorithm that can
be applied to states that are mathematically defined (so any operational
definition of the initial state would eventually have to be converted into a
mathematical one): we may as well work with the mathematical state from the
beginning.

\section{Models of Experiments}

Turn now to the schematic, mathematical descriptions of experiments. Our
assumptions are conventional: we suppose that an experiment is designed to
measure some observable $\widehat{X}$ on a complex Hilbert space $H$, which
for convenience we take to be of finite dimensionality;\footnote{%
It would be just as easy to work with Hilbert spaces of countably inifnite
dimension, and restrict instead the observables to self-adjoint operators
with purely point spectra. (The difficulty with observables with continuous
spectra is purely technical, however.) } we suppose that the apparatus is
prepared in some initial state $\psi ,$ normalized to one,\footnote{%
Later on we shall consider the consequences of relaxing the normalization
condition (correspondingly, we use the term \textquotedblleft state"
loosely, to mean any Hilbert space vector defined up to phase).} and that on
measurement one of a finite number of microscopic outcomes $\lambda _{k}\in
Sp(\widehat{X})$ results, $k=1,...,d$ (we allow for repetitions, i.e. for
some $j\neq k$ we may have $\lambda _{j}=\lambda _{k}).$ We suppose that
these microscopic events are amplified up to the macroscopic level by some
physical process $\Omega :Sp(\widehat{X})\rightarrow U,$ yielding one or
other of the $D$ possible displayed outcomes $u_{j}\in U.$ We suppose the
latter macroscopic events occur with probabilities $p_{j},$ $j=1,..,D$.%
\footnote{%
In the case of the Everett interpretation, we say rather that \textit{all}
of the macroscopic outcomes result, but that each of them is in a different
branch (with a given amplitude). (We will consider the interpretation of
probabilty in the Everett interpretation in due course.)}

We take it that the \textit{details} of the detection and amplification
process are what are disputed, not that there is such a process, nor that it
results in macroscopic outcomes $u_{j}.$ The probabilities computed from
records of repeated trials concern in the first instance these registered,
macroscopic outcomes, not the unobservable microscopic events $\lambda _{k}$
(indeed, on some approaches to foundations, there \textit{are no}
probabilistic microscopic events, prior to amplification up the macroscopic
level). To keep this distinction firmly in mind - and the distinction
between the sets $U$ and $R$ - we shall not assume (as is usual) the
numerical equality of $\Omega (\lambda _{k})$ with $\lambda _{k}$; we do,
however, assume that the macroscopic outcomes $u_{j}\in U$ are physical
numerals, so that addition and multiplication operations can be defined on
them. For convenience we assume that none of these numerals is the \ zero.

Call the triple $\left\langle \psi ,\widehat{X},\Omega \right\rangle $ an 
\textit{experimental model}, denote $g.$ This scheme extends without any
modification to experiments where \ there are inefficiencies in the
detection and registration devices, so long as they are the same for every
channel. (A more sophisticated scheme will be needed if the efficiencies
differ from one channel to the next, however; we neglect this complication
here.)

This scheme applies to a much wider variety of experiments than
multiple-channel experiments; the Born rule is conventionally stated in just
these terms. We shall be interested in algorithms that assign real numbers
to experimental models, interpreted as expectation values, i.e. weighted
averages of the quantities $u_{j},$ with weights given by the probabilities $%
p_{j}$ of each $u_{j},$ $j=1,...,D$. We are therefore looking for a map $%
V:g\rightarrow R$ of the form:

\begin{equation}
V[\psi ,\widehat{X},\Omega ]=\sum_{j=1}^{D}p_{j}u_{j},\text{ }%
\sum_{j=1}^{D}p_{j}=1.
\end{equation}

\noindent If $D=d$ we can write the $u_{j}$'s directly in terms of the $%
\Omega (\lambda _{k})$'s. Otherwise define $\lambda ^{-1}(u_{j})$ $%
=\{k:\Omega (\lambda _{k})=u_{j}\}$, \ $j=1,...,D,$ and choose any real
numbers $w_{k}\in \lbrack 0,1],$ $k=1,...,d$ such that $\sum_{k\in \lambda
^{-1}(u_{j})}w_{k}=p_{j.\text{ }}$From Eq.(1) we obtain:

\begin{equation}
V[\psi ,\widehat{X},\Omega ]=\sum_{k=1}^{d}w_{k}\Omega (\lambda _{k}),\text{ 
}\sum_{k=1}^{d}w_{k}=1.
\end{equation}

\noindent Conversely, given any $d$ real numbers $w_{k}\in \lbrack 0,1]$
satisfying Eq.(2), define the $D$ numbers $p_{j}=\sum_{k\in \lambda
^{-1}(u_{j})}w_{k};$ from Eq.(2) we obtain Eq.(1).

In what follows, we assume the existence of probabilities $p_{j}$ satisfying
Eq.(1), \textit{and therefore} that there are real numbers $w_{k}$
satisfying Eq.(2). The latter will prove more convenient for calculations.

\section{The Consistency Condition}

Our general strategy is as follows. In the special case of multiple-channel
experiments, there are clear criteria for when an experiment is to be
assigned a given model. There follows an important constraint on $V$: for if 
$M$ is assigned two \textit{distinct} models $g,$ $g^{\prime }$, and if
there is to be any general algorithm $V:g\rightarrow R,$ \ then the
expectation values it assigns to these two models had better agree, i.e. $%
V(g)$ $=V(g^{\prime })$. We view this as a \textit{consistency condition} on 
$V.$ Failing this condition, expectation values of models could have no
unequivocal experimental meaning. The probabilistic outcome events $u_{k}\in
U$ that we are talking of are all observable; it is the mean values of these
that the quantities $V(g)$ concern; if one and the same mean value is
matched to two expectation values, $V(g)\neq V(g^{\prime })$, then either
the experiment cannot be modelled by $g$ and $g^{\prime }$, or \textit{there
is no algorithm} $V$ for mapping models to expectation values.

That a condition of this kind played a tacit role in Deutsch's derivation
was recognized by Wallace; it was used explicitly in Wallace's deduction 
\cite{Wallace} of the Born rule, although there it was cast in a slightly
different form, and the conditions for its use were stated in terms of the
Everett theory of measurement (including the theory of the detection and
registration process). Here we make do with operational criteria, and with
assumptions about the behavior of the state \textit{prior }to any detection
events; we suppose that this prior evolution of the state is purely
deterministic, and governed by the unitary formalism of quantum mechanics.%
\footnote{%
Of course in its initial phases the process of state preparation will
involve probabilistic events, if only in collimating particles produced from
the source, or in blocking particular channels. But it does not matter what
these probabilities are; all that matters is that \textit{if }a particle is
located in a given region of the apparatus, \textit{then} it is in a
definite state, and unitarily develops in a definite way (prior to any
detection or registration process).}

Consider a multiple-channel experiment $M.$ By assumption, there are $d$
deterministic experiments $M_{k},$ $k=1,...,d$ that can also be performed
with this apparatus, on blocking every channel save the $k^{\text{th}}$,
each yielding one of the $D$ macroscopic outcomes $u_{j}\in U$. \ Given that
the initial state in region $r$ for $M_{k}$ is $\varphi _{k}$, it is clear
enough, on operational grounds, as to what can be counted as a model for
this experiment: the experiment measures any $\widehat{X}$ such that $%
\widehat{X}\varphi _{k}=\lambda _{k}\varphi _{k},$ for any $\lambda _{k}$
and any $\Omega $ such that $\Omega (\lambda _{k})\in U$ is the outcome of $%
M_{k}.$

Now consider the indeterministic experiment $M$ with every channel open. We
suppose that the state of $M$ at $r$ is $\psi =\sum_{k=1}^{d}c_{k}\varphi
_{k};$ then the observable measured is any $\widehat{X}$ such that $\widehat{%
X}\varphi _{k}=\lambda _{k}\varphi _{k}$ for $k=1,...,d$, and any $\Omega $
such that $\Omega (\lambda _{k})\in U$ \ is the outcome of each $M_{k}$.

Let us state this as a definition:

\begin{definition}
Let $M$ have $d$ channels and $D$ outcomes. Then $M$ realizes $\
\left\langle \psi ,\widehat{X},\Omega \right\rangle $ if and only if

(i) for some region $r$ and orthogonal states $\{\varphi _{k}\}$, $\varphi
_{k}$ is the state of $M_{k}$ in $r$, $k=1,...,d\geq D$, and $\psi
=\sum_{k=1}^{d}c_{k}\varphi _{k}$ is the state of $M$ in $r,$

(ii)$\widehat{X}\varphi _{k}=\lambda _{k}\varphi _{k}$, $k=1,...,d,$

(iii) $\Omega (\lambda _{k})$ is the outcome of $M_{k}$, $k=1,...,d.$
\end{definition}

\noindent The definition applies equally to a deterministic experiment (the
limiting case in which $d=D=1).$ Bearing in mind that from our definition of
multiple-channel experiments, for each $u_{j}\in U$, there is at least one $%
M_{k}$ for which $u_{j}$ is deterministic, it follows from (ii), (iii) that $%
\widehat{X}$ has at least $D$ distinct eigenvalues.

Why is it right to model experiments in this way and not some other? The
deterministic case speaks for itself; in the indeterministic case, the short
answer is that it is underwritten by the linearity of the equations of
motion. An apparatus that \textit{deterministically} measures each
eigenvalue $\lambda _{k}$ of $\widehat{X},$ when the state in a given region
of the apparatus is $\varphi _{k},$ will \textit{indeterministically}
measure the eigenvalues $\lambda _{k}$ of $\widehat{X}$, when the state in
that region is in a superposition of the $\varphi _{k}$'s. This principle is
implicit in standard laboratory procedures; this is how measuring devices
are standardly calibrated, and how their functioning is checked.

The consistency condition now reads:

\begin{definition}
$V$ is consistent if and only if $V(g)=V(g^{\prime })$ whenever $g$ and $%
g^{\prime }$ can be realized by the same experiment.
\end{definition}

In the deterministic case evidently:

\begin{equation}
V[\varphi _{k},\lambda _{k}\widehat{P}\varphi _{k},\Omega ]=\Omega (\lambda
_{k}).
\end{equation}%
We will show that if $|\psi |=1$ and $V$ is consistent, with $\left\langle
,\right\rangle $ the inner product on $H$, then\footnote{%
Whilst $\Omega (\widehat{X})$ makes no sense as an operator (as the values
of $\Omega $ are physical numerals like pointer-positions, not real numbers)
we are assuming that arithmetic operations can be defined for the $\Omega
(\lambda _{k})$'s; define $<\psi ,\Omega (\lambda _{k})\widehat{P}_{\varphi
_{k}}\psi >$ $=\Omega (\lambda _{k})<\psi ,\widehat{P}_{\varphi _{k}}\psi >$
accordingly, and extend by linearity.}

\begin{equation}
V[\psi ,\widehat{X},\Omega ]=\left\langle \psi ,\Omega (\widehat{X})\psi
\right\rangle 
\end{equation}

\noindent Eq.(4) is the Born rule.

We begin with some simple consequences of the consistency condition. The
Born rule is then derived in stages: first for equal norms in the simplest
possible case of a spin half system; then for the general case of equal
norms; and then for rational norms. The general case of irrational norms is
handled by a simple continuity condition. As promised, we shall also derive
a probability rule for initial states normalized to arbitrary finite numbers.

\section{Consequences of the Consistency Condition}

We prove four general constraints on $V$ that follow from consistency.
(Eqs.(5)-(8) may be found in Wallace \cite{Wallace}, derived on somewhat
different assumptions.) In each \ case an equality is derived from the fact
that a single experiment realizes two \textit{different} models: by
consistency, each must be assigned the same expectation value.

We assume it is not in doubt that there do exist such experiments, in which
the initial state (prior to any detection or amplification process) evolves
unitarily in the manner stated.

\begin{lemma}
Let $V$ be consistent. It follows
\end{lemma}

(i) for \bigskip invertible $f:R\rightarrow R$:

\begin{equation}
V[\psi ,\widehat{X},\Omega ]=V[\psi ,f(\widehat{X}),\Omega \circ f^{-1}].
\end{equation}

(ii) For orthogonal projectors $\{\widehat{P}_{k}\},$ $k=1,..,d$, such that $%
\widehat{P}_{k}\varphi _{j}=\delta _{kj}\varphi _{j}$

\begin{equation}
V[\sum_{k=1}^{d}c_{k}\varphi _{k},\sum_{k=1}^{d}\lambda _{k}\widehat{P}%
_{\varphi _{k}},\Omega ]=V[\sum_{k=1}^{d}c_{k}\varphi
_{k},\sum_{k=1}^{d}\lambda _{k}\widehat{P}_{k},\Omega ].
\end{equation}

(iii) For $\widehat{U}_{\theta }:\varphi _{k}\rightarrow e^{i\theta
_{k}}\varphi _{k}$, $k=1,...,d,$ for arbitrary $\theta _{k}\in \lbrack
0,2\pi ]\subset R$

\begin{equation}
V[\psi ,\sum_{k=1}^{d}\lambda _{k}\widehat{P}_{\varphi _{k}},\Omega ]=V[%
\widehat{U}_{\theta }\psi ,\sum_{k=1}^{d}\lambda _{k}\widehat{P}_{\varphi
_{k}},\Omega ].
\end{equation}

(iv) For $\widehat{U}_{\pi }:\varphi _{k}\rightarrow \varphi _{\pi (k)}$,
where $\pi $ is any permutation of $<1,...,d>$%
\begin{equation}
V[\psi ,\widehat{X},\Omega ]=V[\widehat{U}_{\pi }\psi ,\pi ^{-1}(\widehat{X}%
),\Omega ].
\end{equation}

\begin{proof}
Let $g=\left\langle \psi ,\widehat{X},\Omega \right\rangle $ be realized by $%
M$ with $d$ channels. Then for some region $r_{1}$ the state of $M_{k}$ is $%
\varphi _{k},$ $k=1,..,d$, that of $M$ is $\sum_{k=1}^{d}c_{k}\varphi _{k},$
and there exist (not necessarily distinct) real numbers $\lambda
_{1,...,}\lambda _{k}$ such that $\widehat{X}\varphi _{k}=\lambda
_{k}\varphi _{k},$ $\Omega (\{\lambda _{k}\})=U.$ Since for invertible $f$, $%
\Omega \lbrack f^{-1}(f(\lambda _{k})]=\Omega (\lambda _{k}),$ $f(\widehat{X}%
)\varphi _{k}=f(\lambda _{k})\varphi _{k}$, $M$ realizes $\left\langle \psi
,f(\widehat{X}),\Omega \circ f^{-1}\right\rangle $, and (i) follows from
consistency. Further, $M$ realizes any other model $\left\langle \psi ,%
\widehat{Y},\Omega \right\rangle $ $\ $such that $\widehat{Y}\varphi
_{k}=\lambda _{k}\varphi _{k}$; $\sum_{k=1}^{d}\lambda _{k}\widehat{P}_{k}$
is such a $\widehat{Y},$ so (ii) follows from consistency. Suppose now that $%
\psi $ evolves unitarily to the state $\widehat{U}_{\theta }\psi $ in region 
$r_{2}$. Then in $r_{2}$ the state of each $M_{k}$ is $e^{i\theta
_{k}}\varphi _{k}$, and since $\widehat{P}_{e^{i\theta _{k}}\varphi _{k}}=%
\widehat{P}_{\varphi _{k}}$, $M$ realizes $\left\langle \widehat{U}_{\theta
}\psi ,\sum_{k=1}^{d}\lambda _{k}\widehat{P}_{\varphi _{k}},\Omega
\right\rangle $, and (iii) follows from consistency. Finally, let $\psi $
subsequently evolve to the state $\widehat{U}_{\pi }\psi $ in region $r_{3}$%
. Then in $r_{3}$ the state of each $M_{k}$ is $\varphi _{\pi (k)},$ and the
state of $M$ \ is $\sum_{k=1}^{d}c_{k}\varphi _{\pi (k)}$. Without loss of
generality, we may write $\widehat{X}$ as $\sum_{k=1}^{d}\lambda _{k}$ $%
\widehat{P}_{\varphi _{k}}$; then $\pi ^{-1}(\widehat{X})=\sum_{k=1}^{d}%
\lambda _{k}$ $\widehat{P}_{\varphi _{\pi (k)}}$ satisfies $\pi ^{-1}(%
\widehat{X})\varphi _{\pi (k)}=\lambda _{k}\varphi _{\pi (k)}$, so $M$
realizes $\left\langle \widehat{U}_{\pi }\psi ,\pi ^{-1}(\widehat{X}),\Omega
\right\rangle $, and (iv) follows from consistency
\end{proof}

Eqs.(5)-(8) are of course trivial consequences of the Born rule, Eq.(4).
Note further that in each case the observables whose expectation values are
identified \textit{commute} - these are constraints among probability
assignments to projectors belonging to a \textit{single} resolution of the
identity. Finally, note that the normalization of the initial state $\psi $
played no role in the proofs.

\section{Case 1: The Stern-Gerlach Experiment for Equal Norms}

Consider the Stern-Gerlach experiment with $d=D=2.$ Let $\widehat{X}=$ $%
{\frac12}%
\widehat{P}_{+}-%
{\frac12}%
\widehat{P}_{-}=\widehat{\sigma }_{z}$ (in conventional notation), the
observable for the $z$-component of spin with eigenstates $\varphi _{\pm }$,
and let $\psi =c_{+}\varphi _{+}+c_{-}\varphi _{-}$. Let $\widehat{U}_{\pi }$
interchange $\varphi _{+}$ and $\varphi _{-}$, so $\widehat{U}_{\pi }%
\widehat{\sigma }_{z}\widehat{U}_{\pi }^{-1}=-\widehat{\sigma }_{z}.$ From
Lemma 3(iv) it follows that:

\begin{equation}
V[c_{+}\varphi _{+}+c_{-}\varphi _{-},\widehat{\sigma }_{z},\Omega
]=V[c_{+}\varphi _{-}+c_{-}\varphi _{+},-\widehat{\sigma }_{z},\Omega ].
\end{equation}

\noindent From Eq.(9) and Lemma 3(i):

\begin{equation}
V[c_{+}\varphi _{+}+c_{-}\varphi _{-},\widehat{\sigma }_{z},\Omega
]=V[c_{+}\varphi _{-}+c_{-}\varphi _{+},\widehat{\sigma }_{z},\Omega \circ
-I]
\end{equation}

\noindent (where $(\Omega \circ -I)(x)=\Omega (-x)$)$.$ From Eq.(10), in the
special case that $|c_{+}|^{2}=|c_{-}|^{2},$ and using Lemma \ 3(iii) to
compensate for any differences in phase:

\begin{equation}
V[c_{+}\varphi _{+}+c_{-}\varphi _{-},\widehat{\sigma }_{z},\Omega
]=V[c_{+}\varphi _{+}+c_{-}\varphi _{-},\widehat{\sigma }_{z},\Omega \circ
-I].
\end{equation}%
Consider the LHS of this equality. From Eq.(2), writing $w_{1}=w$, $%
w_{2}=1-w,$ $\Omega (\pm 
{\frac12}%
)=$ $\Omega (\pm )$ - so that $\Omega (+)$ results with probability $w,$ and 
$\Omega (-)$ results with probability $1-w)$ - we obtain the expectation
value $x=w\Omega (+)+(1-w)\Omega (-).$ But by similar reasoning, the RHS
yields $w\Omega (-)+(1-w)\Omega (+)=-x+\Omega (+)+\Omega (-).$ Equating the
two, $x=%
{\frac12}%
[\Omega (+)+\Omega (-)].$

We have shown, for $|c_{+}|^{2}=|c_{-}|^{2}:$

\begin{equation}
V[c_{+}\varphi _{+}+c_{-}\varphi _{-},\widehat{\sigma }_{z},\Omega ]=%
{\frac12}%
\Omega (+)+%
{\frac12}%
\Omega (-)
\end{equation}%
in accordance with the Born rule. Note that here we have derived an
expectation values in a situation (dimension 2) where Gleason's theorem does
not apply. (Note that the normalization of the initial state $\psi $ is
again irrelevant to the result.)

\section{Case 2: General Superpositions of Equal Norms}

Consider an arbitrary observable on any $d-$dimensional subspace $H_{d}$ of
Hilbert space. By the spectral theorem, we may write $\widehat{X}%
=\sum_{k=1}^{d}\lambda _{k}\widehat{P}_{\varphi _{k}},$ for some set of
orthogonal vectors $\{\varphi _{k}\}$, $k=1,...,d$ spanning $H_{d}$, where
there may be repetitions among the $\lambda _{k}$'s. Let $\psi $ be a
(not-necessarily normalized) vector in $H_{d}$; then for some $d$-tuple of
complex numbers $<c_{1},...,c_{d}>$, $\psi =\sum_{k=1}^{d}c_{k}\varphi _{k}.$
For any permutation $\pi $, we have from Lemma 3(iv), (i):

\begin{equation}
V[\sum_{k=1}^{d}c_{k}\varphi _{k},\widehat{X},\Omega
]=V[\sum_{k=1}^{d}c_{k}\varphi _{\pi (k)},\pi ^{-1}(\widehat{X}),\Omega
]=V[\sum_{k=1}^{d}c_{k}\varphi _{\pi (k)},\widehat{X},\Omega \circ \pi ].
\end{equation}

\noindent

\noindent If $|c_{k}|^{2}=|c_{j}|^{2},$ $j,k=1,...,d$, using Lemma 3(iii) as
before to adjust for any phase differences

\begin{equation}
V[\psi ,\widehat{X},\Omega ]=V[\psi ,\widehat{X},\Omega \circ \pi ].
\end{equation}%
Let $<w_{1},...,w_{d}>$ be a $d$-tuple of non-negative real numbers
satisfying Eq.(2). From Eq.(14):

\begin{equation}
\sum_{k=1}^{d}w_{k}\Omega (\lambda _{k})=\sum_{k=1}^{d}w_{k}\Omega (\lambda
_{\pi (k)}).
\end{equation}%
Eq.(15) holds for any permutation; let $\pi $ interchange $j$ and $k$, and
otherwise act as the identity. There follows

\begin{equation}
w_{j}\Omega (\lambda _{j})+w_{k}\Omega (\lambda _{k})=w_{k}\Omega (\lambda
_{j})+w_{j}\Omega (\lambda _{k}).
\end{equation}%
Conclude that if $\Omega (\lambda _{j})\neq \Omega (\lambda _{k})$ then $%
w_{k}=w_{j}$ (recall that by convention $0\notin U$, so $\Omega (\lambda
_{k})$ is never zero). 

If $D=d,$ evidently $w_{k}=w_{j}$ for all $j$,$k=1,...,d$. Since $%
\sum_{k}^{d}w_{k}=1$, $\ w_{k}=\frac{1}{d},$ $k=1,...,d.$ Therefore

\begin{equation}
V[\psi ,\widehat{X},\Omega ]=\frac{1}{d}[\sum_{k=1}^{d}\Omega (\lambda _{k}).
\end{equation}

\noindent If not, suppose $\Omega (\lambda _{j})=\Omega (\lambda _{k})$ for $%
j,k=1,...,b<d.$ (If $b=d$ Eq.(17) follows trivially.) For any $j$, $k$ such
that $b<j\leq d,$ $k\leq b,$ $\Omega (\lambda _{k})\neq $ $\Omega (\lambda
_{j})$, from which we conclude as before that $w_{k}=w_{j}.$ Note further
that under the stated conditions, $1/d=|c_{k}|^{2}(%
\sum_{j=1}^{d}|c_{j}|^{2})^{-1}.$ We have proved

\begin{theorem}
Let $\psi =\sum_{k=1}^{d}c_{k}\varphi _{k},$ where $|c_{k}|^{2}=|c_{j}|^{2}$
for all $j,k=1,...,d.$ Then if $V$ is consistent%
\begin{equation}
V[\sum_{k=1}^{d}c_{k}\varphi _{k},\sum_{k=1}^{d}\lambda _{k}\Omega _{\varphi
_{k}},\Omega ]=\sum_{k=1}^{d}\frac{|c_{k}|^{2}}{\sum_{j=1}^{d}|c_{j}|^{2}}%
\Omega (\lambda _{k}).
\end{equation}
\end{theorem}

Like Lemma 3, Theorem 4 is independent of the normalization of $\psi .$

\section{Case 3: d=2 Normalized Superpositions with Rational Norms}

The idea for extending these methods to treat the case of unequal but
rational norms is as follows: consider an experiment in which the initial
state $\psi $ evolves deterministically so that each component $\varphi _{k}$
entering into the initial superposition with amplitude $c_{k}$ evolves into
a superposition of $z_{k}$ orthogonal states of equal norm $1/\sqrt{z_{k}}$,
such that $|c_{k}/\sqrt{z_{k}}!^{2}$ is constant for all $k$. One can then
show that the experiment has a model in which the initial state is a
superposition of states of equal norms, so Theorem 4 can be applied.
(Evidently for this to work each $|c_{k}|^{2}$ will have to be a rational
number.)

For simplicity, consider first the case $d=2$ for real amplitudes. Let $\psi
=\frac{\sqrt{m}}{\sqrt{m+n}}\varphi _{1}+\frac{\sqrt{n}}{\sqrt{m+n}}\varphi
_{2}$, where $m$ and $n$ are integers. Let $\widehat{X}=\lambda _{1}\widehat{%
P}_{\varphi _{1}}+\lambda _{2}\widehat{P}_{\varphi _{2}}$. We will show that
if $V$ is consistent, $V[\psi ,\widehat{X},\Omega ]=\frac{m}{m+n}\Omega
(\lambda _{1})+\frac{n}{m+n}\Omega (\lambda _{2}).$Let the deterministic
experiments of $M$ be $M_{1},M_{2},$ with registered outcomes $\Omega
(\lambda _{1}),$ $\Omega (\lambda _{2})$ respectively. Let the initial \
states of $M,$ $M_{1},M_{2}$ in region $r_{1}$ be $\psi ,$ $\varphi
_{1},\varphi _{2}$ respectively. Then $M$ realizes $g_{1}=\left\langle \psi ,%
\widehat{X},\Omega \right\rangle .$ Now let $\psi $ evolve to $\widehat{U}%
\psi $ in region $r_{2},$ where $\widehat{U}\varphi _{1}=\frac{1}{\sqrt{m}}%
\sum_{k=1}^{m}\chi _{k}$, $\widehat{U}\varphi _{2}=\frac{1}{\sqrt{n}}%
\sum_{k=m+1}^{n+m}\chi _{k}$, for some orthogonal set of vectors $\{\chi
_{k}\}$, $k=1,...,m+n.$ Denote $\lambda _{1}\widehat{P}_{\widehat{U}\varphi
_{1}}+\lambda _{2}\widehat{P}_{\widehat{U}\varphi _{2}}$ by $\widehat{Y}.$
Then the initial state of $M_{i}$, $i=1,2$ is $\widehat{U}\varphi _{i}$ in $%
r_{2},$ whilst that of $M$ is $c_{1}\widehat{U}\varphi _{1}+c_{2}\widehat{U}%
\varphi _{2}$ in $r_{2}$; since $\widehat{Y}\widehat{U}\varphi _{i}=\lambda
_{i}\widehat{U}\varphi _{i}$ it follows that $M$ realizes $%
g_{2}=\left\langle \widehat{U}\psi ,\widehat{Y},\Omega \right\rangle $. By
consistency, $V(g_{1})=V(g_{2}).$ Now define $\widehat{P}_{1}=\lambda
_{1}\sum_{k=1}^{m}\widehat{P}_{\chi _{k}},$ $\widehat{P}_{2}=\lambda
_{2}\sum_{k=m+1}^{n+m}\widehat{P}_{\chi _{k}}$; since $\widehat{P}_{k}%
\widehat{U}\varphi _{j}=\delta _{kj}\widehat{U}\varphi _{j}$, $k,j=1,2$, by
Lemma 3(ii) it follows $V[\widehat{U}\psi ,\widehat{Y},\Omega ]=V[\widehat{U}%
\psi ,\lambda _{1}\widehat{P}_{1}+\lambda _{2}\widehat{P}_{2},\Omega ].$ But 
$\widehat{U}\psi =\frac{1}{\sqrt{m+n}}\sum_{k=1}^{n+m}\chi _{k}$ ; applying
Theorem 4 for $d=m+n$, and noting that $\Omega (\lambda _{k})=\lambda _{1}$
for $k=1,...,m$, and $\lambda _{2}$ otherwise, the result follows.

\section{Case 4: General Superpositions with Rational Norms}

The argument just given assumed $\psi $ was normalized to one. The standard
rational for this is of course based on the probabilistic interpretation of
the state, and hence, at least tacitly, on the Born rule. It may be objected
that we are only able to derive the dependence of the expectation value on
the \textit{squares} of the norms of the initial state, because this is put
in by hand from the \ beginning. But this suspicion is unfounded. Suppose,
indeed, only that $\frac{|c_{1}|^{2}}{|c_{2}|^{2}}$ $=\frac{m}{n}.$ As
before, define $\widehat{U}\varphi _{1}=\frac{1}{\sqrt{m}}\sum_{k=1}^{m}\chi
_{k}$, $\widehat{U}\varphi _{2}=\frac{1}{\sqrt{n}}\sum_{k=m+1}^{n+m}\chi
_{k}.$The state $\widehat{U}\psi $ in region $r_{2}$ will have whatever
normalization $\psi $ had in $r_{1}$; the states $\widehat{U}\varphi _{i}$ $%
i=1,2$ will be eigenstates of $\widehat{P}_{i}$, as before; Definition 1,2
will apply as before. Conclude that if $V$ is consistent, $V[\psi ,\widehat{X%
},\Omega ]=$ $V[\widehat{U}\psi ,\lambda _{1}\widehat{P}_{1}+\lambda _{2}%
\widehat{P}_{2},\Omega ]$, as before. The difference is that now $\widehat{U}%
\psi =\frac{c_{1}}{\sqrt{m}}\sum_{k=1}^{m}\chi _{k}+\frac{c_{2}}{\sqrt{n}}%
\sum_{k=m}^{n+m}\chi _{k}=\frac{c_{1}}{\sqrt{m}}\sum_{k=1}^{n+m}\chi _{k}$ $=%
\frac{c_{2}}{\sqrt{n}}\sum_{k=1}^{n+m}\chi _{k}$ (adjusting the phases of $%
c_{1}$ and $c_{2}$ , using \ Lemma 3(ii), as required). Evidently we have an
initial state which is a superposition of \ $n+m$ components of equal norm, $%
m$ of which \ yield outcome $\Omega (\lambda _{1})$ and $n$ of which yield
outcome $\Omega (\lambda _{2}).$ Since $\frac{m}{n+m}=\frac{|c_{1}|^{2}}{%
|c_{1}|^{2}+|c_{2}|^{2}},$ $\frac{n}{n+m}=\frac{|c_{2}|^{2}}{%
|c_{1}|^{2}+|c_{2}|^{2}}$

\begin{equation}
V[\psi ,\lambda _{1}\widehat{P}_{\varphi _{1}}+\lambda _{2}\widehat{P}%
_{\varphi _{2}},\Omega ]=\frac{|c_{1}|^{2}}{|c_{1}|^{2}+|c_{2}|^{2}}[\Omega
(\lambda _{1})]+\frac{|c_{2}|^{2}}{|c_{1}|^{2}+|c_{2}|^{2}}[\Omega (\lambda
_{2})].
\end{equation}

\noindent Evidently the normalization of $\psi $ is irrelevant.

This result is worth proving in full generality:

\begin{theorem}
For each $i,j=1,...,d$ let $c_{i}\in C$ satisfy $|c_{i}|>0,$ $\frac{%
|c_{i}|^{2}}{|c_{j}|^{2}}\in Z$. Then $\ $%
\begin{equation}
V[\sum_{k=1}^{d}c_{k}\varphi _{k},\sum_{k=1}^{d}\lambda _{k}\widehat{P}%
_{\varphi _{k}},\Omega ]=\sum_{k=1}^{d}\frac{|c_{k}|^{2}}{%
\sum_{j=1}^{d}|c_{j}|^{2}}\Omega (\lambda _{k}).
\end{equation}

\begin{proof}
For $\{c_{k}\}$ as stated, there exists $c\in C$, $z_{k}\in Z,$ $\theta
_{k}\in \lbrack 0,2\pi ],$ $k=1,..,d$ such that $c_{k}=ce^{i\theta _{k}}%
\sqrt{z_{k}}.$ Let $m_{k},n$ be integers such that $z_{k}=\frac{m_{k}}{n},$ $%
k=1,...,d$; let $\{\chi _{j}\},$ $j=1,...,s$ be an orthonormal basis on an $%
s-$dimensional subspace of Hilbert space $H_{s}$, where\ $%
s=\sum_{s=1}^{d}m_{s}$ (we may suppose for $j=1,...,d,$ $\chi _{j}=\varphi
_{j}).$ Define $\widehat{U}$ on $H^{s}$ by the action $\widehat{U}\varphi
_{k}=\frac{1}{\sqrt{m_{k}}}\sum_{j=m_{k}+1}^{m_{k+1}}\chi _{j}$; let $%
\widehat{P}_{k}=\widehat{P}_{\widehat{U}\varphi _{k}},$ $k=1,...,d.$ Let $%
\psi =$ $\sum_{k=1}^{d}c_{k}\varphi _{k};$ let $M$ realize $%
g_{1}=\left\langle \psi ,\sum_{k=1}^{d}\lambda _{k}\widehat{P}_{\varphi
_{k}},\Omega \right\rangle $; then for some region $r_{1},$ the initial
state of $M$ is $\psi $ and the state of each $M_{k}$ is $\varphi _{k}$ with
outcome $\Omega (\lambda _{k}).$ Let the state of $M$ at $r_{2}$ be $%
\widehat{U}\psi $; then $M$ also realizes $g_{2}=\left\langle \widehat{U}%
\psi ,\sum_{k=1}^{d}\lambda _{k}\widehat{P}_{k},\Omega \right\rangle $, and
by consistency $V(g_{1})=V(g_{2}).$ But by construction%
\begin{equation}
\widehat{U}\psi =\sum_{k=1}^{d}c_{k}\widehat{U}\varphi _{k}=\sum_{k=1}^{d}%
\frac{c_{k}}{\sqrt{m_{k}}}\sum_{j=m_{k}+1}^{m_{k+1}}\chi _{j}=\sum_{k=1}^{d}%
\frac{ce^{i\theta _{k}}}{n}\sum_{j=m_{k}+1}^{m_{k+1}}\chi _{j}
\end{equation}%
so $V[\widehat{U}\psi ,\sum_{k=1}^{d}\lambda _{k}\sum_{s=m_{k}+1}^{m_{k+1}}%
\widehat{P}_{\chi _{s}},\Omega ]=V[\frac{c}{n}\sum_{k=1}^{s}\chi
_{s},\sum_{k=1}^{d}\lambda _{k}\sum_{s=m_{k}+1}^{m_{k+1}}\widehat{P}_{\chi
_{s}}\Omega ]$ (by Lemma 3(ii)). The result follows from Theorem 4 (of $s$
equiprobable outcomes, $m_{k}$ have outcome $\Omega (\lambda _{k}),$ so $%
V(g)=\frac{1}{s}\sum_{k=1}^{d}m_{k}\Omega (\lambda _{k}).$ But $%
m_{k}/s=m_{k}(\sum_{j=1}^{d}m_{j})^{-1}=z_{k}(%
\sum_{j=1}^{d}z_{j})^{-1}=|c_{k}|^{2}\sum_{j=1}^{d}|c_{j}|^{2})^{-1})$
\end{proof}
\end{theorem}

Examination of the proof shows that the dependence of probabilities on the
modulus square of the expansion coefficients of the state ultimately derives
from the fact that we are concerned with unitary evolutions on Hilbert
space, specifically an \textit{inner-product }space, and not some general
normed linear topological space. A general class of norms on the latter is
of the form $\left\Vert x\right\Vert =\left( \sum_{k=1}^{d}|\xi
_{k}|^{p}\right) ^{1/p}$, $1\leq p<\infty $ ($d$ may also be taken as
infinite). Such spaces ($l^{p}$ spaces) are metric spaces and can be
completed in norm. The proof as we have developed it would apply equally to
a theory of unitary (i.e. invertible norm-preserving) motions on such a
space, yielding the probability rule

\begin{equation}
V[\sum_{k=1}^{d}c_{k}\varphi _{k},\sum_{k=1}^{d}\lambda _{k}\widehat{P}%
_{\varphi _{k}},\Omega ]=\sum_{k=1}^{d}\frac{|c_{k}|^{p}}{%
\sum_{j=1}^{d}|c_{j}|^{p}}\Omega (\lambda _{k})
\end{equation}

\noindent (assuming that $\frac{|c_{i}|^{p}}{|c_{j}|^{p}}\in Z$, $%
j,k=1,...,d)$. But the only space of this form which is an inner-product
space is $p=2$ (Hilbert space). 

\section{Case 5: Arbitrary States}

There are a variety of possible strategies for the treatment of irrational
norms, but the one that is most natural, given that we are making use of
operational criteria for the interpretation of experiments, is to weaken
these criteria in the light of the limitations of \textit{realistic}
experiments. In practise, one would not expect \textit{precisely} the same
state to be prepared on each run of the experiment. Properly speaking, the
statistics actually obtained will be those for an \textit{ensemble} of
experiments; correspondingly, they should be obtained from a family of
models, differing slightly in their initial states. We should therefore
speak of\textit{\ approximate} models (or of models that are \textit{%
approximately} realized) - where the differences among the models are small.

How small is small? What is the topology on the space of states? The obvious
answer, from a theoretical point of view, is the norm topology. We should
suppose that for sufficiently small $\epsilon $, so long as $|\psi -\psi
^{\prime }|<\epsilon $, then if $\left\langle \psi ,\widehat{X},\Omega
\right\rangle $ is an approximate model for $M$ then so is $\left\langle
\psi ^{\prime },\widehat{X},\Omega \right\rangle $. Indeed, $\widehat{X}$
and $\Omega $ will likewise be subject to small variations. (Only the
outcome set $U$ can be regarded as precisely specified, insofar as outcomes
are identified with numerals.)

But now it is clear that the details are hardly important; any algorithm
that applies to families of models of this type, yielding expectation
values, will have to be \textit{continuous} in the norm topology. Given
that, the extension of Theorem 5 to the irrational case is trivial. We
define:

\begin{definition}
Let $g^{(i)}$ be any sequence of models $\left\langle \psi ^{(i)},\widehat{X}%
,\Omega \right\rangle $, $i=1,2,...$such that $\underset{i\rightarrow \infty 
}{\lim }|\psi ^{(i)}-\psi |=0.$ Then $V$ is continuous in norm if $\underset{%
i\rightarrow \infty }{\lim }V($ $g^{(i)})=V(g).$
\end{definition}

We may finally prove:

\begin{theorem}
Let $V$ be consistent and continuous in norm. Then for any model $%
\left\langle \psi ,\widehat{X},\Omega \right\rangle $%
\begin{equation}
V[\psi ,\widehat{X},\Omega ]=\frac{\left\langle \psi ,\Omega (\widehat{X}%
)\psi \right\rangle }{\left\langle \psi ,\psi \right\rangle }.
\end{equation}
\end{theorem}

\begin{proof}
It is enough to prove that any realizable model satisfies Eq.(22). If
realizable, there is some multiple-channel experiment $M$ with $d$ channels
and $D$ outcomes that realizes $\left\langle \psi ,\widehat{X},\Omega
\right\rangle $. Let $\{\varphi _{k}\}$, $\ k=1,..,d$ be any orthogonal
family of vectors such that $\widehat{X}\varphi _{k}=\lambda _{k}\varphi _{k}
$ (not all the $\lambda _{k}$'s need be distinct). Without loss of
generality, let $\psi =\sum_{k=1}^{d}c_{k}\varphi _{k},$ $\widehat{X}=$ $%
\sum_{k=1}^{d}\lambda _{k}\widehat{P}_{\varphi _{k}}.$ Let $0<\epsilon
=\sum_{k=1}^{d}|c_{k}|^{2}$ and let $\{c_{k}^{(i)}\}\sqsubseteq C^{d}$ be
any sequence of $d$-tuples such that $\epsilon \leq
\sum_{k=1}^{d}|c_{k}^{(i)}|^{2},\frac{|c_{j}^{(i)}|^{2}}{|c_{k}^{(i)}|^{2}}$ 
$\subset Z$, $\underset{i\rightarrow \infty }{\lim }c_{k}^{(i)}=c_{k}$ (such
a sequence can always be found). Let $\psi ^{(i)}=$ $%
\sum_{k=1}^{d}c_{k}^{(i)}\varphi _{k}$, $g^{(i)}=\left\langle \psi ^{(i)},%
\widehat{X},\Omega \right\rangle $. By Theorem 5, $V[\psi ^{(i)},\widehat{X}%
,\Omega ]=\sum_{k=1}^{d}\frac{|c_{k}^{(i)}|^{2}}{%
\sum_{j=1}^{d}|c_{j}^{(i)}|^{2}}\Omega (\lambda _{k})=\frac{1}{%
\sum_{j=1}^{d}|c_{j}^{(i)}|^{2}}\sum_{k=1}^{d}\Omega (\lambda
_{k})\left\langle \psi ^{(i)},\widehat{P}_{\varphi _{k}}\psi
^{(i)}\right\rangle .$ The numerator is $\left\langle \psi
^{(i)},\sum_{k=1}^{d}\Omega (\lambda _{k})\widehat{P}_{\varphi _{k}})\psi
^{(i)}\right\rangle $ (by the continuity of the inner product), i.e. $%
\left\langle \psi ^{(i)},\Omega (\widehat{X})\psi ^{(i)}\right\rangle ;$
since the denominator is bounded below by $\epsilon >0,$ with $\underset{%
i\rightarrow \infty }{\lim }\sum_{j=1}^{d}|c_{j}^{(i)}|^{2}=%
\sum_{j=1}^{d}|c_{j}|^{2}$, and since $\underset{i\rightarrow \infty }{\lim }%
\left\langle \psi ^{(i)},\Omega (\widehat{X})\psi ^{(i)}\right\rangle
=\left\langle \psi ,\Omega (\widehat{X})\psi \right\rangle $ (again by the
continuity of the inner product), the result follows from the continuity of $%
V$
\end{proof}

\noindent A similar proof can be given for a general probability rule on $%
l^{p}$ spaces, $p\neq 2$ (i.e. Eq.(22), for arbitrary complex coefficients;
of course this result could not be expressed as in Eq.(23), using an inner
product). 

Is a continuity assumption permitted in the present context? Gleason's
theorem does not require it; if one is going to do better than Gleason's
theorem, it would be pleasant to derive the continuity of the probability
measure, rather than to assume it. But from an operational point of view
continuity is a very natural assumption: no algorithm that could ever be
used is going to distinguish between states that differ infinitesimally.

\section{A Role for Decision Theory}

Deutsch \cite{Deutsch} took a rather different view: he was at pains to
establish the Born rule for irrational norms, without assuming continuity.
His method, however, was far from operational: along with axioms of decision
theory, he assumed that quantum mechanics is \textit{true} (under the
Everett interpretation).

A hybrid is possible: the present method can in fact be supplemented with
axioms of decision theory, yielding the Born rule for irrational norms,
without any continuity assumption. But as Wallace \cite{Wallace}\ makes
clear, nothing much hangs on this question. One can do without a continuity
assumption, but there are just as good reasons to invoke it from a decision
theoretic point of view as from an operational one. In neither case is there
any reason to distinguish between states that differ infinitesimally.

Decision theory is important for a rather different reason: it is because
the non-probabilistic parts of decision theory (as Deutsch puts it), or
decision theory in the face of uncertainty (as Wallace puts it) \textit{can
provide an account of probability in terms of something else}. This matters
in the case of the Everett interpretation; according to many, the Everett
interpretation has no place for probability \cite{Kent}; given Everett,
probability cannot be taken as \textit{primitive}.

So it is clear why Deutsch took the more austere line: if Everett is to be
believed, quantum mechanics \textit{is }purely deterministic. Deutsch
supposed that the fundamental concept (that can be taken as primitive) is
rather the \textit{value} or the \textit{utility }that an agent places upon
a model - that $V(g)$ is in fact a utility. He argued that experiments
should be thought of as games; for each registered outcome in $U,$ we are to
associate some utility, fixed in advance. So, in effect, the mapping $\Omega
:\lambda _{k}\rightarrow \Omega (\lambda _{k})\in U$ defines the \textit{%
payoff }for the outcome $\lambda _{k}.$

Decision theory on this approach has a substantial role. If we suppose that
the utilities of a rational agent are ordered, and satisfy very general
assumptions (\textquotedblleft axioms of rationality"), a representation
theorem can be derived \cite{Savage} which \textit{defines} subjective
probability in terms of the ordering of an agent's utilities. In effect, one
deduces - in accordance with these axioms - that the agent acts \textit{as
if }she places such-and-such subjective probabilities on the outcomes of
various actions.\footnote{%
This does not mean that subjective probabilities are illusory, and
correspond to nothing in reality. The point is to \textit{legitimate} the
concept, not to abolish it. As for its objective correlate, the most popular
candidate has long been relative frequency (of outcomes in a sequence of
trials). Relative frequencies are obviously important when it comes to 
\textit{evidence} for probabilities, but there are well-known difficulties
with trying to\textit{\ identify them} with probabilities (for anything
short of infinite sequences). We read Everett as making a contrary proposal:
that the objective correlates of subjective probability are branches in the
universal state (with respect to the decoherence basis). Here we are
deducing the quantitative rule to be used in assigning subjective
probabilities to branches.}

It is important that one can still make sense of uncertainty in this
context, as Wallace explains. It may be we cannot help ourselves to
probabilistic ideas \textit{ab initio}, but that does not mean that one only
deals with certainties - that games, in some sense, have only a \textit{%
single }payoff, as Deutsch at one point suggests \cite[p.3132-3.]{Deutsch}
From a first-person perspective, one does not know what outcome of a quantum
game to expect to observe (there is certainly no first-person perspective
from which they can \textit{all} be observed). In fact, it is enough that -
in the face of branching - a rational agent expects anything at all (that
she does not expect \textit{oblivion} \cite{Saunders}).

On this line of thought, the proofs of the Born rule just presented make an
illegitimate assumption: Eq.(1). We are not entitled to assume that the
macroscopic outcomes $u_{j}\in U,$ $j=1,...,D$ occur with probabilities $%
p_{j}$, for they all occur; so neither can we assume there are non-negative
real numbers, summing to one, satisfying Eq.(2). But the proof of Theorem 4
(hence 5 and 7) depended on this assumption. \ Of course we may, with
Deutsch and Wallace, eventually be in a position to make statements about
the subjective probabilities of branches, but if so such statements will
have to come at a later stage - \textit{after }establishing the values $V(g)$
of various games. But then how are we to establish these values?

Here Wallace has provided a considerably more detailed analysis than
Deutsch, and from weaker premises. But the proofs are correspondingly more
complicated; for the sake of simplicity we shall only consider Deutsch's
argument, removing the ambiguities of notation in the way shown by Wallace.

First, consider Case 1, the Stern-Gerlach experiment. All is in order up to
Eq.(11), but we must do without the assumption subsequently made - that the
registered outcome $\Omega (+)$ results with probability $w,$ and outcome $%
\Omega (-)$ with probability $1-w.$ Here Deutsch invokes a new principle,
what he calls \textit{the zero-sum rule}:

\begin{equation}
V[\varphi ,\widehat{X},\Omega ]=-V[\varphi ,\widehat{X},-\Omega ].
\end{equation}%
Following Deutsch, let us assume that the numerical value of the utility $%
\Omega (\lambda _{k})$ equals $\lambda _{k}.$ Then, in the special case
where $\lambda _{1}=-\lambda _{2}$ (true for the measurement of a component
of spin), from Eq.(24), applied to Eq.(11), we deduce:

\begin{equation}
V[c_{1}\varphi _{1}+c_{2}\varphi _{2},\widehat{\sigma }_{z},\Omega
]=-V[c_{1}\varphi _{1}+c_{2}\varphi _{2},\widehat{\sigma }_{z},\Omega ]
\end{equation}%
and hence that $\ V[c_{1}\varphi _{1}+c_{2}\varphi _{2},\widehat{\sigma }%
_{z},\Omega ]=0,$ in accordance with the Born rule in this special case.

Although evidently of limited generality, the result is illustrative -
assuming the zero-sum rule can be independently justified. (Of course it
follows trivially from Eq.(2), but this was derived from Eq.(1), and at this
point we cannot make use of the concept of probability.) Here is an
argument: banking too is a form of gambling; the only difference between
acting as the gambler who bets, and as the banker who accepts the bet, is
that whereas the gambler \textit{pays} a stake in order to play, and \textit{%
receives} payoffs according to the outcomes, the banker \textit{receives }%
the stake in order to play, and \textit{pays} the payoffs according to the
outcomes. The zero-sum rule is the statement that the most that one will pay
in the hope of gaining a utility is the least that one will accept to take
the risk of losing it. We may take it that \textit{this} principle, as a
principle of zero-sum games, is perfectly secure. And \ evidently any
quantum experiment can be used to play a zero-sum game; therefore this
principle also applies to the \ expected utility of experiments.

What of the general equal-norm case, Case 2? Here the zero-sum rule is not
enough. But if we consider only the case $d=2$, it is enough to supplement
it with another rule, what Deutsch calls the \textit{additivity} rule. A
payoff function $\Omega :R\rightarrow U$ is \textit{additive} if and only if 
$\Omega (x+y)=\Omega (x)+\Omega (y)$. Let $f_{k}:R\rightarrow R$ be the
function $f_{k}(x)=x+k$; then $V$ is \textit{additive }if and only if

\begin{equation}
V[\psi ,\widehat{X},\Omega \circ f_{k}]=V[\psi ,\widehat{X},\Omega ]+\Omega
(k).
\end{equation}

\noindent Additivity of the payoff function is a standard assumption of
elementary decision theory, eminently valid for small bets (but hardly valid
for large ones, or for utilities that only work in tandem). Additivity of $V$
then has a clear rational: it is an example of a \textit{sure-thing principle%
}, that if, given two games, each exactly the same, except that in one of
them one receives an additional utility $\Omega (k)$ \textit{whatever} the
outcome, then one should value that game as having an \textit{additional }%
utility $\Omega (k).$

To see how additivity can be used in Case 2 (but restricted to $d=2$),
observe that for $k=-\lambda _{1}-\lambda _{2},$ the function $-I\circ f_{k}$
is the permutation $\pi .$ Therefore from Eq.(14) we may conclude:

\begin{equation}
V[\psi ,\widehat{X},\Omega ]=V[\psi ,\widehat{X},\Omega \circ -I\circ f_{k}].
\end{equation}%
By additivity the RHS is $V[\psi ,\widehat{X},\Omega \circ -I]+(\Omega \circ
-I)(k)$, and since $\Omega $ is additive (so $\Omega \circ -I=-\Omega $) we
obtain, from the zero-sum rule

\begin{equation}
V[\psi ,\widehat{X},\Omega ]=-V[\psi ,\widehat{X},\Omega ]-\Omega (k).
\end{equation}

\noindent With a further application of payoff additivity there follows

\begin{equation}
V[\psi ,\widehat{X},\Omega ]=%
{\frac12}%
[\Omega (\lambda _{1})+\Omega (\lambda _{2})]
\end{equation}%
in accordance with the Born rule.

As Wallace has shown, this, along with the higher dimensional cases ($d>2$),
can be derived from much weaker axioms of decision theory, that do not
assume additivity. \ Theorem 5 then goes through unchanged.\footnote{%
It is worth remarking that a derivation of the Born rule for initial states
that are not normalized to unity is just what is needed for the Everett
interpretation, as also the de Broglie-Bohm theory (in reality, according to
either approach, one \textit{always} deals with branch amplitudes with
modulus strictly less than one - supposing the initial state of the universe
has modulus one).} As already remarked, one is then in a position to derive
the extension to the irrational case without assuming continuity: for the
details, I refer to Wallace \cite{Wallace}.

Decision theory can evidently play a role in the derivation of the Born
rule, but it is only needed if the notion of probability is itself in need
of justification. That may well be so, in the context of the Everett
interpretation; but on other approaches to quantum mechanics, probability,
whatever it is, can be taken as given.

\section{Gleason's Theorem}

Compare Gleason's theorem:

\begin{theorem}
Let $f$ be any function from 1-dimensional projections on a Hilbert space of
dimension $d>2$ to the unit interval, such that for each resolution of the
identity $\{\widehat{P}_{k}\},$ $k=1,...,d$, $\sum_{k=1}^{d}\widehat{P}%
_{k}=I $, $\sum_{k=1}^{d}f(\widehat{P}_{k})=1.$ Then there exists a unique
density matrix $\rho $ such that $f(\widehat{P}_{k})=Tr(\rho \widehat{P}%
_{k}).$
\end{theorem}

\begin{proof}
Gleason (1967)
\end{proof}

A first point is that the derivation of the Born rule presented here
concerns the notion of a \textit{fixed} algorithm that applies to arbitrary
measurement models, hence to Hilbert spaces of \textit{arbitrary} dimension,
whereas Gleason's theorem concerns an algorithm that applies to \textit{%
arbitrary }resolutions of the identity on a Hilbert space of \textit{fixed}
dimension$.$ Although the proof of Theorems 5 and 7 made use of a Hilbert
space of large dimensionality, it applies to the 2-dimensional case as well.

More important, on a variety of approaches to quantum mechanics, nothing so
strong as Gleason's premise is really motivated. It is not required that \
probabilities can be defined for a projector independent of the family of
projectors of which it is a member. This requirement, sometimes called 
\textit{non-contextuality }\cite{Redhead}, is very strong. Very few
approaches to quantum mechanics subscribe to it. The theorem has no
relevance to any approach that singles out a unique basis once and for all:
it applies neither to the GRW theory \cite{Ghirardi}, nor to the de
Broglie-Bohm theory \cite{Holland}, which single out the position basis; it
does not apply to the Everett interpretation \cite{de Witt}, which singles
out a basis approximately localized in phase space; it does not apply to the
consistent histories approach \cite{Dowker}, assuming the choice of
decoherent history space is unique. All these theories require \textit{only}
that probabilities be defined for projectors associated with the preferred
basis - if they apply to any other resolution of the identity, it is insofar
as in a \textit{particular} context, experimental or otherwise, the latter
projections become correlated with members of the former family.

But so much is entirely compatible with the derivation that we have offered.
By all means restrict Definition 1 to observables compatible with a unique
resolution of the identity (and likewise the consistency condition of
Definition 2). Lemma 3 proves identities for expectation values for
commuting observables, it likewise can be restricted to a unique resolution
of the identity; likewise Theorem 4. In Theorem 5 an auxiliary basis was
used, but again this can again be taken as the preferred basis. And whilst
it is In the spirit of Theorem 7 that probabilities should also be defined
for small variations in projectors, this does not yet amount to the
assumption of non-contextuality.

Unlike the premise of Gleason's theorem, the operational criteria that we
have used are hardly disputed; they are common ground to all the major
schools of foundations of quantum mechanics. But it would be wrong to
suggest that they apply to all of them equally: on some approaches - in
particular, those that provide a detailed dynamical model of measurements -
there is good reason to suppose that an algorithm for expectation values
will depend on additional factors (in particular, on the state at the
instant of state reduction); the Born rule may no longer be forced in
consequence. (But we take it that this would \ be an \textit{unwelcome}
consequence of these approaches; the \ Born rule will have to be \textit{%
otherwise} justified - presumably, as it is in the GRW theory, as a \textit{%
hypothesis}).

Of the major schools, two - the Everett interpretation, and those based on
operational assumptions (here we include the Copenhagen interpretation) -
offer no such resources. This point is clear enough in the latter case; in
the case of the Everett interpretation, the association of models with
multiple channel experiments as given in Definition 1 follows from the full
theory of measurement.\footnote{%
For arguments in the still more general case, on applying the Everett theory
of measurement \ to any experiment, I refer to Wallace \cite{Wallace} (see
in particular his principle of \textquotedblleft measurement neutrality").}
Quantum mechanics under the Everett interpretation provides no leeway in
this matter. The same is likely to be true of any approach to quantum
mechanics that preserves the unitary formalism intact, without any
supplement to it.

The principal remaining schools have a rather different status. One, the
state-reduction approach, has already been remarked on: a new and detailed
dynamical theory of measurement is likely to offer novel definitions of
experimental models and novel criteria for when they are to be applied. The
other is the hidden-variable approach, in which the state evolves unitarily
even during measurements (but is incomplete). This case deserves special
consideration.

\section{Completeness}

As it happens, the one approach to foundations in which the Born rule has
been seriously questioned is an example of this type (the de Broglie-Bohm
theory)\cite{Valentini}. Hidden variables certainly make a difference to the
argument we have presented. Consider the proof of Theorem 4. The passage
from Eq.(13) to Eq.(14) hinged on the fact that the state on both sides of
Eq.(13) is \textit{identical }when the norms of its components are the same.
(Likewise the step from Eq.(10) to (11).) \ But if the state is incomplete,
this is not enough to ensure the required identification. Including the
state of the hidden variables as well (denote $w$), we should replace $\psi $
by the pair $<\psi ,\omega >$ ($w$ may be the value of the hidden variable,
or a probability distribution over its values). \ Doing this, as Wallace has
pointed out \cite{Wallace}, there is no guarantee that in the case of
superpositions of equal norms - e.g. for $\psi =\frac{1}{\sqrt{2}}(\varphi
_{1}+\varphi _{2}),$ where $\varphi _{1},$ $\varphi _{2}$ are, as in Case 1,
eigenstates of the $z-$component of spin - that $\widehat{U}_{\pi }$
(permuting $\varphi _{1}$ and $\varphi _{2})$ will act as the identity.
Although $\widehat{U}_{\pi }\psi =\psi $, its action on $<\psi ,\omega >$
may well be different from the identity; how is the permutation to act on
the hidden variables?

The question is clearer when $\widehat{U}_{\pi }$ implements a spatial
transformation. We have an example where it does: the Stern-Gerlach
experiment. In this case $\widehat{U}_{\pi }\widehat{\sigma }_{z}\widehat{U}%
_{\pi }^{-1}=-\widehat{\sigma }_{z},$ a reflection in the $x-y$ plane. Under
the latter, a particle initially with positive $z$-coordinate ($\omega =+$)
is mapped to one with negative $z$-coordinate ($\omega =-$). Under this same
transformation, the superposition $\psi =\frac{1}{\sqrt{2}}(\varphi
_{1}+\varphi _{2})$ is unchanged. Therefore $\widehat{U}_{\pi }:<\psi
,+>\rightarrow <\psi ,->\neq <\psi ,+>$; there is no longer any reason to
suppose that Eq.(11) will be satisfied.

This situation is entirely as expected. In the de Broglie-Bohm theory, given
such an initial state $\psi ,$ it is well known that if the incident
particle is located on one side of the plane of symmetry of the
Stern-Gerlach apparatus, then it will always remain there. It is obvious
that if the particles is always located on the same side of this plane, on
repetition of the experiment, the statistics of the outcomes will disagree
with the Born rule. It is equally clear that if particles are randomly
distributed about this plane of symmetry then the Born rule will be obeyed -
but that is only to say that the probability distribution for the hidden
variables is determined by the state, in accordance with the Born rule. This
is what we are trying to prove.

But it does not follow that the arguments we have given have no bearing on
such a theory. Our strategy, recall, was to derive constraints on an
algorithm - any algorithm - that takes as inputs experimental models and
yields as outputs expectation values. The constraints will apply even if the
state is incomplete, even if there are additional parameters controlling
individual measurement outcomes - so long as the state \textit{alone}
determines the statistical distribution of the hidden variables. Given that,
then any symmetries of the state will also be symmetries of the distribution
of hidden variables. In application to the de Broglie-Bohm theory, our
result indeed implies that the particle distribution \textit{must} be given
by the Born rule - this is no longer an additional postulate of the theory - 
\textit{so long as the particle distribution is determined only by the state}%
. The assumption is not that particles must be distributed in accordance
with the Born rule, but that they are distributed by any rule at all that is
determined by the state$.\bigskip $ Then it is the Born rule.

\noindent \textbf{Acknowledgements}{\LARGE . }The author would like to thank
Harvey Brown, and especially David Wallace, for detailed discussions.


\begin{thebibliography}{99}
\bibitem{Barnum} Barnum, H., C. Caves, J. Finkelstein, C. Fuchs, and R.
Schack, \textquotedblleft Quantum Probability from Decision Theory?",
Proceedings of the Royal Society of London A456 1175-1182, (2000); available
online at http://xxxx.arXiv.org/abs/quant-ph/9907024.

\bibitem{Deutsch} Deutsch, D., \textquotedblleft Quantum Theory of
Probability and Decisions", \textit{Proceedings of the Royal Society of
London }A455 3129-3137, (1999); available online at
http://xxxx.arXiv.org/abs/quant-ph/9906015.

\bibitem{Dowker} Dowker, H.F. and J.J. Halliwell: \textquotedblleft Quantum
Mechanics of History: The Decoherence Functional in Quantum
Mechanics\textquotedblright , \textit{Physical Review, D46}, 1580- 609
(1992).

\bibitem{Gleason} Gleason, \ A., \textquotedblleft Measures on the Closed
Subspaces of a Hilbert space", \textit{Journal of Mathematics and Mechanics}%
, 6, 885-894, (1967).

\bibitem{Ghirardi} Ghirardi, G.C., A. Rimini and T.Weber,\textit{\ Physical
Review}, D34, 470. (1986)

\bibitem{Holland} Holland, P., \textit{The Quantum Theory of Motion},
Cambridge: Cambridge University Press (1993).

\bibitem{Kent} Kent, A. \textquotedblleft Against Many-Worlds
Interpretations", \textit{International Journal of Modern Physics, A5},
1745-62, (1990).

\bibitem{Redhead} Redhead, M., \textit{Incompleteness, Nonlocality, and
Realism}, Oxford (1987).

\bibitem{Saunders} Saunders, S., \textquotedblleft Time, Quantum Mechanics,
and Probability", \textit{Synthese}, 114, 373-404; available online at
http://xxxx.arXiv.org/abs/quant-ph/01

\bibitem{Savage} Savage, L., \textit{The Foundations of Statistics}, 2$^{%
\text{nd}}$ Ed., New York, Dover (1972).

\bibitem{Valentini} Valentini, A., \textquotedblleft Hidden Varianbles,
Statistical Mechanics, and the Early Universe", available online at
http://xxxx.arXiv.org/abs/quant-ph/0104067.

\bibitem{Wallace} Wallace, D. \textquotedblleft Quantum Probability and
Decision Theory, Revisited", available online at
http://xxxx.arXiv.org/abs/quant-ph/0211104.
\end{thebibliography}
\end{document}